\documentclass[ 
    aps,
    prl,
    twocolumn,
    letterpaper,
    10pt,
    superscriptaddress,
    showpacs,
    showkeys,
    notitlepage,
    amsmath, 
    amssymb, 
    floatfix 
]{revtex4} 

\usepackage[english]{babel}
\usepackage{indentfirst}
\usepackage{graphicx}
\usepackage{amsmath}
\usepackage{amssymb}
\usepackage{amsthm}
\usepackage{mathrsfs}
\usepackage[linktocpage={true}]{hyperref}
\usepackage{bbold}
\usepackage{lscape}
\usepackage{bm}
\usepackage{bbm}
\usepackage{float}
\usepackage{longtable}
\usepackage{color}
\usepackage[utf8x]{inputenc}

\input{Qcircuit}

\def\ket#1{\left|#1\right>}
\def\bra#1{\left<#1\right|}

\def\E#1{\langle #1 \rangle}
\def\eqref#1{Eq.~(\ref{#1})}
\def\figref#1{Fig.~\ref{#1}}

\newcommand{\ketbra}[2]{\vert #1\rangle \! \langle #2\vert}
\newcommand{\G}{\mathcal G}

\def\dblone{\hbox{$1\hskip -1.2pt\vrule depth 0pt height 1.6ex width 0.7pt
\vrule depth 0pt height 0.3pt width 0.12em$}}

\begin{document}
\title{Quantum Error Correction for Metrology}

\author{E. M. Kessler}
\affiliation{Physics Department, Harvard University, Cambridge,
MA 02138, USA}
\affiliation{ITAMP, Harvard-Smithsonian Center for Astrophysics, Cambridge, MA 02138, USA}

\author{I. Lovchinsky}
\affiliation{Physics Department, Harvard University, Cambridge,
MA 02138, USA}

\author{A. O. Sushkov}
\affiliation{Physics Department, Harvard University, Cambridge,
MA 02138, USA}
\affiliation{Department of Chemistry and Chemical Biology, Harvard University, Cambridge,
MA 02138, USA}

\author{M. D. Lukin}
\affiliation{Physics Department, Harvard University, Cambridge,
MA 02138, USA}


\vspace{-3.5cm}

\date{\today}
\begin{abstract}
We propose and analyze a new approach based on quantum error correction (QEC) to improve quantum metrology in the presence of noise. We identify the conditions under which QEC allows one to improve the signal-to-noise ratio in quantum-limited measurements, and we demonstrate that it enables, in certain situations, Heisenberg-limited sensitivity.
 We discuss specific applications to nanoscale sensing using nitrogen-vacancy centers in diamond in which QEC can significantly improve the measurement sensitivity and bandwidth under realistic experimental conditions.
\end{abstract}
 \maketitle

The precise measurement of physical quantities is of great importance in science, with implications ranging from the \textit{global positioning system} (GPS), to nanoscale sensing in biological systems \cite{Maze:2008ch}, and tests of fundamental laws of physics \cite{Schiller2008,Wolf2008}. The theory of quantum metrology \cite{Helstrom1969,Giovannetti2006} provides an efficient framework for understanding the fundamental limits of the achievable accuracy in the determination of a parameter (e.g., a magnetic field or frequency), given a certain amount of resources (e.g., number of available atoms or time). In recent years, the exploration of these limits in the presence of realistic imperfections and noise have been actively pursued \cite{Huelga1997,Escher:2011fn,Koiodynski:2013bu,DemkowiczDobrzanski:2012gl,Anonymous:GuhK9eRJ}.
In a typical quantum measurement, the sensing qubits (e.g., atoms or spins) repeatedly interrogate the parameter to be measured over a total time $\tau$.
For instance, a Ramsey-type experiment involves a sequence of measurement cycles of duration $T$. 
Since each Ramsey cycle introduces measurement noise it is beneficial to extend the duration of a single interrogation to its maximum value $T\rightarrow \tau$. However, in the presence of qubit decoherence of rate $\gamma$, the Ramsey time is inherently limited, as for times $T\geq 1/\gamma$ the phase information acquired during the interrogation is lost \cite{Huelga1997}.   
One of the most successful techniques to counter the compromising effects of environmental noise is \textit{dynamical decoupling} (DD) \cite{Viola:1999jw,Uhrig:2007ek} which has become a standard technique, e.g., in coherent solid state physics to increase qubit lifetimes for quantum information processing \cite{Maurer:2012kg}. Here, a series of control pulses (or continuous wave control fields) effectively achieves a cancellation of the coupling Hamiltonian between the system (i.e., qubit) and its environment to a certain order, thus effectively reducing the value of $\gamma$ \cite{Viola:1999jw}. 
DD can also enhance the sensitivity in quantum metrology \cite{Taylor2008}. Recently, it has been successfully used to improve the signal-to-noise ratio in magnetometry \cite{Maze:2008ch,London:2013fq,Mamin:2013eu,Staudacher:2013kn}, and temperature measurement \cite{Kucsko:2013gq}.
However, in order to achieve sensitivity improvements, the pulse repetition rate of a DD protocol (which has to match the frequency of the measured signal) needs to be faster than the correlation time of the environmental bath $\tau_c$. Therefore,  for environments with fast internal dynamics DD is not feasible.


In this Letter, we propose a complementary approach that employs quantum error correction (QEC) \cite{Shor:1995fj,Gottesman:2009ug,Nielsen:2000vn} to enhance the qubit coherence for metrology. 
In contrast to DD, the QEC operations have to be implemented on timescales of the \textit{error rate} $\gamma$. The effectiveness of our approach is therefore independent of the correlation time of the bath, and it is capable of correcting noise even in the limit of Markovian environments ($\tau_c\rightarrow0$). 
Our protocol can be applied to improve metrology with individual qubits. The most direct application is to nanoscale measurements of magnetic and electric fields using nitrogen vacancy (NV) centers in diamond (relevant, e.g., for studies of neural activity to magnetic imaging of biomolecules and exotic materials). We show that in such measurements significant improvements in sensitivity and detector bandwidth can be obtained. Our approach can be understood as a sequential feedback protocol. When applied to ensembles of $N$ qubits, it can yield, in certain situations, Heisenberg-limited scaling, thus surpassing the recently developed sensitivity bounds in the presence of noise \cite{Anonymous:GuhK9eRJ,Koiodynski:2013bu,DemkowiczDobrzanski:2012gl}.

QEC is based on the fact that any kind of noise, discrete or continuous, can be represented in  Kraus decomposition by a discrete set of error operation elements $\{E_0,\dots E_w\}$. It is then possible -- by the use of redundant degrees of freedom (provided, e.g., by ancilla qubits) -- to encode the logical information in a 
subspace of the Hilbert space (the so called \textit{quantum code} $\mathcal C$) such that each of the errors $E_i$ maps the code to an respective orthogonal and undeformed subspace $\mathcal E_i$, allowing to efficiently detect and correct whenever an error has occurred.
The challenge in QEC for metrology is to devise a code which is capable to reliably identify and correct errors while, at the same time, not interfering with the signal. 
Consider a state that evolves within the code space $\mathcal C$ under the action of the Hamiltonian $H$ generating the signal we aim to measure: $e^{-iHt}\ket{\Psi} = \ket{\Psi_{\phi(t)}}\in \mathcal C$ for $\ket{\Psi}\in \mathcal C$  ($\phi(t)$ denotes a parametrization of the state evolution, e.g., by the phase accumulated in a Ramsey-type experiment). If an error $E_i$ occurs the state is mapped to $E_i  \ket{\Psi_{\phi(t)}} = \ket{\eta_{\phi(t)}} \in \mathcal E_i$. In the simplest case, the spaces $ \mathcal E_i$ and $ \mathcal C$ are orthogonal, and we are able to reliably detect this error by measurement of the projector on $ \mathcal E_i$ (the so-called \textit{syndrome operator}). Evidently, this is not always possible, e.g., in  the case where the generator of the signal is proportional to the error operation element $H\propto E_i$. Any conceivable QEC code will also "correct" the signal, and compromise the sensitivity of the detecting state. 
In what follows, we derive a general set of conditions under which QEC can be employed to improve metrology.


\textit{General Formalism.} We consider a generic scenario in quantum metrology. 
Let us assume we have $N$ detector qubits to sense a parameter $\omega$, e.g., a magnetic or electric field.
Interrogating the parameter, the qubits evolve coherently according to the Hamiltonian
\begin{align}
\label{eq:ham}
H_s = \frac{\omega}{2} \G,
\end{align}
where $\G$ represents the generator of the signal which, in general, can be a sum of single- or multi-particle operators. During the evolution, the qubits are subject to some arbitrary form of noise which is described by the quantum operation \cite{Nielsen:2000vn}
\begin{align}
\mathcal E (\rho) = \sum_k E_k \rho E_k^\dagger.
\end{align}
 If we further denote the completely positive map corresponding to the coherent evolution of \eqref{eq:ham} as $\mathcal M \rho = e^{-i H t}\rho e^{i H t}$, the goal of QEC for metrology is to design a \textit{recovery operation} $\mathcal R$, such that
 \begin{align}
\label{eqn:QEC}
(\mathcal R  \circ \mathcal E\circ \mathcal M) (\rho) \propto  \mathcal M \rho,
\end{align}
for all states within a certain \textit{quantum code} $\rho \in \mathcal C \leq \mathcal H$, where $\mathcal H$ denotes the Hilbert space. Note, that \eqref{eqn:QEC} has to be understood in the short-time limit where $\mathcal E\circ \mathcal M \approx \mathcal M\circ \mathcal E$, i.e., recovery operations have to be applied on timescales short compared to the noise rate $\gamma$. 
Defining $P$ as the projector on the code space $\mathcal C$, the recovery operation $\mathcal R$ of \eqref{eqn:QEC} exists iff the two conditions
\begin{enumerate}
\item $[\G,P]=0$,
\item $PE_i^\dagger E_j P =A_{i,j}P$,
\end{enumerate}
are fulfilled, with $A=(A_{i,j})$ being a hermitian matrix. Condition (1) represents the requirement that the code $\mathcal C$ is an invariant subspace of the generator $\G$, and ensures that if we prepare a code state and no error occurs, the evolution is restricted to the code space: $\mathcal M(\rho) \in\mathcal C,~ \forall \rho\in\mathcal C$. Condition (2) guarantees that the error operation elements $E_i$ map the code space onto orthogonal and undeformed subspaces, and in a constructive proof \cite{Nielsen:2000vn} one can show that conditions (1) \& (2) guarantee the existence of the recovery operation defined in \eqref{eqn:QEC} which is able to correct the errors $E_i$, without disturbing the signal.

However, these conditions alone allow also for solutions, in which the generator $\G$ acts as the identity on the code. Obviously, such a code is useless for metrology, since the action of the Hamiltonian yields a global phase on the code states. In order to exclude these trivial solutions we further require that the maximum 
quantum Fisher information \cite{Giovannetti2011} within the code space must be larger than zero
\begin{enumerate}
   \setcounter{enumi}{2}
\item $\xi\equiv\underset{\ket\Psi\in\mathcal C}{\textrm{max}} \left< \Delta \G^2\right>_\Psi>0$,
\end{enumerate}
where $ \left< \Delta \G^2\right>_\Psi=\bra{\Psi} \mathcal G^2 \ket{\Psi} - \bra{\Psi} \mathcal G\ket{\Psi}^2$. Since the achievable precision in the measurement of $\omega$ is $\delta \omega \propto 1/\sqrt\xi$ \cite{Giovannetti2006}, $\xi$ also serves as a figure of merit
which quantifies how useful a particular code $\mathcal C$ is for metrology.

%

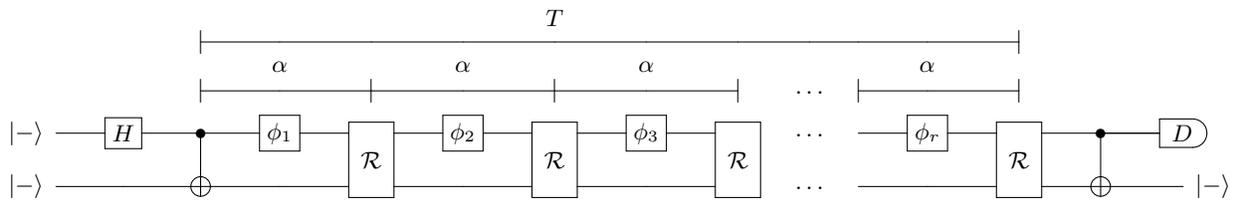
\begin{figure*}[t]
\label{fig:circuit1}$
\begin{matrix}
  \Qcircuit @C=2em @R=1em {  &&&&&&&T&&&&&&&\\
   &                                             &           &            \mid  &             \qw       &               \qw         &      \qw                 &               \qw                &          \qw         &        \qw            &             \qw&       \qw           &    \qw      &                           \qw           \mid    &\\
  &&&&\alpha&&\alpha&&\alpha&&&&\alpha&&\\
  &                                             &           &            \mid  &             \qw       &               \qw           \mid &      \qw                 &               \qw           \mid        &          \qw         &        \qw           \mid         &      \hdots       &        \mid  &             \qw       &               \qw           \mid    &                              &\\
&\lstick{\ket -}          &\gate H& \ctrl{1}      & \gate{\phi_1}   &\multigate{1}{\mathcal R}& \gate{\phi_2}      &\multigate{1}{\mathcal R} &  \gate{\phi_3}&\multigate{1}{\mathcal R}&\hdots&&\gate{\phi_r}&\multigate{1}{\mathcal R}&\ctrl{1}&\measureD{D} \qw&\\ 
 & \lstick{\ket -}        & \qw     & \targ         & \qw                  & \ghost{\mathcal R}     & \qw                        & \ghost{\mathcal R} & \qw&\ghost{\mathcal R}& \hdots& &\qw& \ghost{\mathcal R}&\targ&\rstick{\ket{-}}\qw&
 }
 \end{matrix}$
 \caption{Circuit model of the error correction for the model described by \eqref{eq:ham2}. The code state can be prepared by application of a Hadamard (denoted by $H$) and CNOT gate. 
 After each segment of free evolution of duration $\alpha = T/r$ the QEC operation $\mathcal R$ is applied. After the final decoding and measurement ($D$) the effective error rate has been reduced by a factor of the number of QEC steps $r$.}
 \end{figure*}

\textit{Example.} To illustrate the general working principle, consider the following model system of a single qubit detector subject to phase-flip noise in $z$ direction (\textit{pure dephasing}) sensing a signal in $x$ direction described by the Hamiltonian
\begin{align}
\label{eq:ham2}
H_{s} = \frac{\omega}{2}X_1,
\end{align}
where $X_1$ is the $x$ Pauli operator acting on the detector spin ($Z_1$ and $Y_1$, respectively, will denote the remaining Pauli matrices below). The noise is described by the operation elements $E_0=\sqrt{1- p} \dblone, ~E_1 =\sqrt p Z_1$, where $p$ is the error probability.
Using standard Ramsey spectroscopy, the qubit interrogates the parameter for the Ramsey time $T$, and after $n$ repetitions we can determine the value of $\omega$ with accuracy \cite{Itano1993}
\begin{align}
\delta \omega \approx \frac{1}{T \sqrt n}=\frac{1}{\sqrt{T\tau}},
\end{align}
where we defined the total measurement time $\tau = nT$. Due to the presence of noise, the Ramsey time is limited by the dephasing rate $T\leq 1/\gamma$ ($\Leftrightarrow p=\gamma T\leq 1$), resulting in the suboptimal measurement accuracy
\begin{align}
\delta \omega \leq \sqrt \frac{\gamma}{\tau}.
\end{align}

Let us now assume we additionally have a long-lived and  "blind" ancilla at our disposal which neither interacts with the parameter nor is subject to noise. By defining the simple code spanned by the two states $\ket{1}\equiv \ket{++}$ and  $\ket{0}\equiv \ket{--}$ (where $\pm$ in the first (second) slot represents $X$ eigenstates of the single detector (ancilla) qubit), one readily checks that, $[H_s,P]=0$, $ \xi = 2$, and $A= \text{diag}(1-p, p)$, i.e. the requirements for QEC are met. To perform the measurement, we initialize the system in the state $\ket\Psi=(\ket{++} +\ket{--})/\sqrt2 \in\mathcal C$. Under the action of the Hamiltonian the state accumulates a phase $\phi = \omega t$: $\ket{\Psi(t)}\propto (\ket{++} + e^{-i \phi}\ket{--})/\sqrt2$. If a $Z$ error occurs the state is mapped to $\ket{\Psi(t)}\propto (\ket{-+} + e^{-i \phi}\ket{+-})/\sqrt2$, such that the subsequent evolution reduces the phase, rather than increasing it, resulting in a randomized signal for $T\geq \gamma^{-1}$

\begin{figure}[!]
\centering
\includegraphics[width=0.45\textwidth]{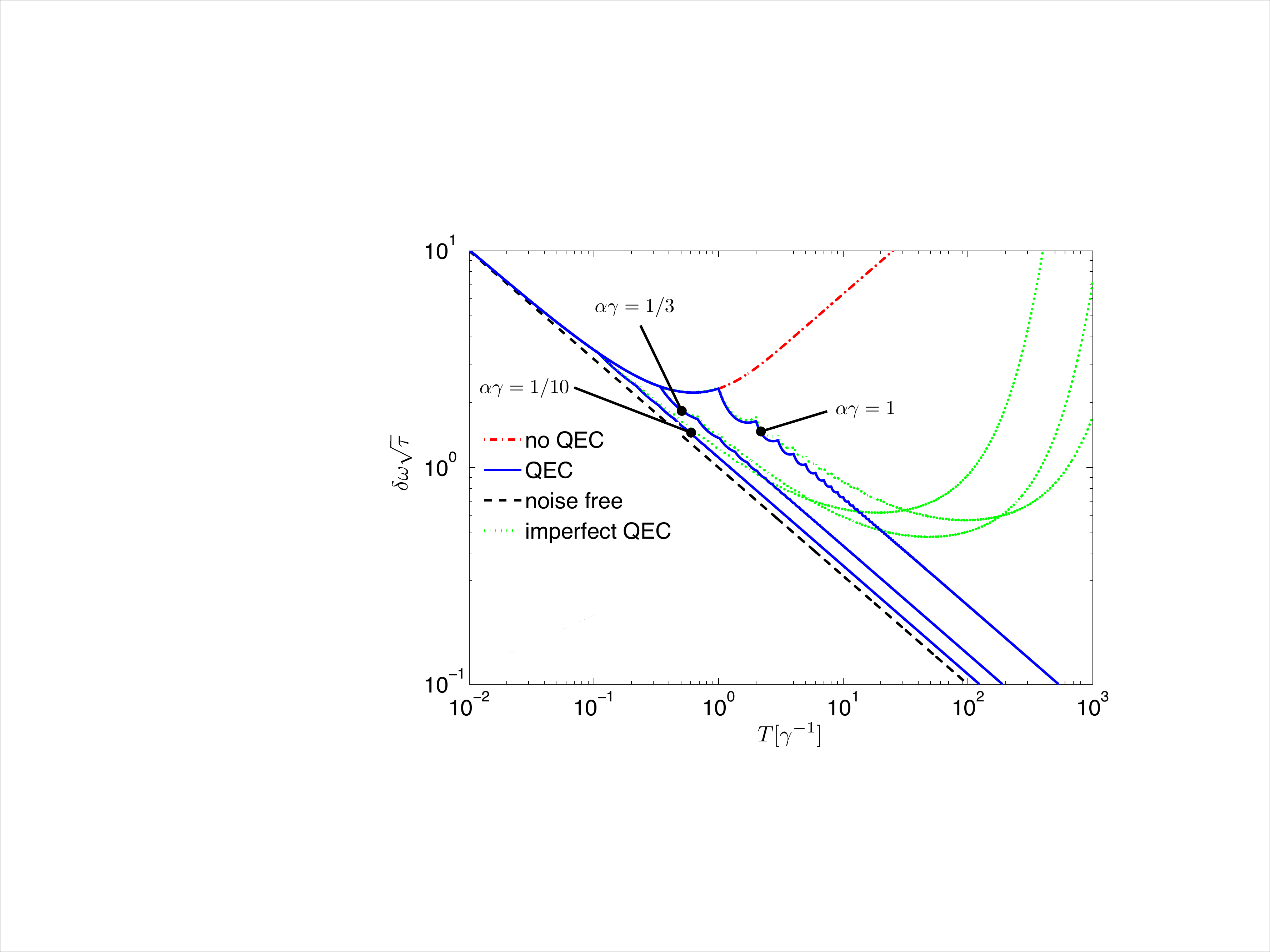}
\caption{ Estimation error $\delta\omega$ for the model of transversal noise (see text), normalized by the total measurement time $\tau$ (for sufficiently large $\tau$). In the standard approach (dash-dotted) the free evolution time has an optimal value $T\approx \gamma^{-1}$, limiting the achievable sensitivity. Ideally, QEC (solid) can restore the noise-free scaling (dashed) with the interrogation time ($\propto 1/\sqrt T$) even for relatively small QEC repetition rates $1/\alpha = \gamma$. The dotted lines show the achievable sensitivity in the presence of a
 small parallel noise component $\gamma_{\parallel}=10^{-3}\gamma$, and a probability $p_\text{error}=10^{-3}$ that the QEC operation fails. 
\label{fig:2} }
\end{figure}

To implement QEC, we divide the Ramsey time $T$ into $r$ intervals of equal duration $\alpha = T/r$, and perform a QEC step $\mathcal R$ after each segment ($\mathcal R $ is assumed to be instantaneous on timescales of the evolution), as illustrated in \figref{fig:circuit1}1. The QEC operation $\mathcal R$ consists of two steps: 
\cite{fn11}:
\begin{enumerate}
\item Measuring the syndrome operator $X_1 X_2$ (with $X_2$ acting on the ancilla spin).
\item For outcome $-1$:  Application of an $X_1$ gate. \\For outcome $+1$:  No action is required.
\end{enumerate}
Although single errors within a segment can be corrected with the operation $\mathcal R$ (assuming perfect gates), they introduce a small phase uncertainty, due to the fact that the exact time of the error within the interval $\alpha$ is unknown. 
Despite this small residual uncertainty, we demonstrate in \cite{SI} that by performing $r$ QEC steps we can extend the Ramsey time linearly to a value $T \rightarrow r\gamma^{-1}$, assuming the QEC operation $\mathcal R$ is implemented on a timescale short compared to the dephasing time $\gamma^{-1}$.
Consequently, after  $r\approx\gamma \tau^{-1}$ repetitions we can extend the interrogation time to its maximum value $T \rightarrow \tau$, and achieve the best sensitivity allowed by quantum mechanics 
\begin{align}
\label{eq:result}
\delta\omega \approx  1/\tau.
\end{align}
This result is confirmed by numerical simulations displayed in \figref{fig:2}. Even for relatively low repetition rates of the recovery operations, $\alpha \gamma =1$, the linear, noise-free scaling is recovered. For imperfect recovery operations (failing with probability $p_\text{error} = 10^{-3}$), and residual parallel noise components ($\gamma_\parallel = 10^{-3} \gamma$), a significant constant improvement is found. 

Quantum metrology in the presence of perpendicular noise as described by \eqref{eq:ham2} has been investigated in \cite{Anonymous:GuhK9eRJ} for the case of multi-particle measurements. Evaluating the general precision bounds derived in \cite{Koiodynski:2013bu,DemkowiczDobrzanski:2012gl} yields an optimal asymptotic scaling of the sensitivity $\delta\omega \propto  1/(N^{5/6}\sqrt \tau)$. While this result represents a scaling better than the standard quantum limit (i.e., $\propto 1/\sqrt N$), it can further be improved by allowing for sequential feedback protocols, as represented by the QEC-based method we now suggest. 
Being provided with $N$ detector spins we define the code  $\ket{1}\equiv \ket{+}^{\otimes{N}}$ and  $\ket{0}\equiv \ket{-}^{\otimes N}$ (note that here no ancilla is needed). Assuming independent $Z$ noise acting on the individual detector spins, the error operation elements are given as $E_0 = \sqrt{1-N p} \dblone$, and $E_i = p Z_i$ (i=1\dots N), where we neglect operation elements of order $\mathcal O (p^2)$ or higher. 
Again, one readily checks that all requirements for QEC are fulfilled with $\xi= (2N)^2$, indicating the potential for Heisenberg-limited spectroscopy. We prepare the system in a Greenberger-Horne-Zeilinger (GHZ) state $\ket{\Psi} =(\ket{+}^{\otimes{N}} + \ket{-}^{\otimes{N}})/\sqrt2 \in \mathcal C$, which accumulates the phase $\Phi$ $N$ times faster than uncorrelated qubits. In this situation, a single error $Z_i$ can be detected by measuring the syndrome operators $X_{i-1}X_i$ and  $X_{i}X_{i+1}$, and corrected by an appropriate $\pi$ rotation. A single QEC operation $\mathcal R$ consequently involves $N-1$ syndrome measurements of the operators $X_iX_i+1$ (i=1\dots N-1). As above, repetitive application of $\mathcal R$ allows to extend the Ramsey time to the maximum value $T\rightarrow\tau$, achieving, in principle, the Heisenberg limit of metrology \cite{Giovannetti2006} $\delta \omega \approx 1/(N\tau)$ with an optimal scaling in both resources time $\tau$ and particle number $N$ \cite{fn1}.

These considerations demonstrate that under certain conditions QEC (and possibly other feedback protocols) provide a way to improve the sensitivity bounds derived in \cite{Anonymous:GuhK9eRJ,Koiodynski:2013bu,DemkowiczDobrzanski:2012gl}. We note that this result does not contradict \cite{Escher:2011fn}, where it was demonstrated that feedback strategies cannot improve sensitivity in phase estimation if the associated channel is full rank. In our protocol the QEC operation is employed explicitly before the channel associated to the free system evolution under the particular noise model we consider becomes full rank, i.e. in the short-time limit.

\textit{Applications.} 
The feasibility of quantum QEC has recently been demonstrated experimentally in various different physical systems, such as trapped ions \cite{Schindler:2011ch}, superconducting qubits \cite{Reed:2012dz}, and nitrogen-vacancy (NV) centers \cite{Waldherr:2013uq,Taminiau:2013up}.
In the following, we consider an example from solid state nano-sensing using nitrogen-vacancy (NV) defect centers in diamond in which our approach can be applied under realistic experimental conditions. Recent work \cite{Dolde:2011hi} has suggested and experimentally demonstrated the use of NV centers for sensing of electric fields with high sensitivity and spatial resolution, e.g., for the biological imaging of neural activity \cite{Peterka:2011js,Robinson:2012fw,Kralj:2011hv,Alivisatos:2013ek,Alivisatos:2012iu}. NV centers are optically addressable diamond lattice defects with a stable paramagnetic ground state of spin $S=1$ \cite{Jelezko:2006}. In zero magnetic field, the spin state $\ket 0$ is separated from the degenerate states $\ket{1}$ and $\ket{-1}$ by a splitting of $\omega_0\sim3$~GHz. 
Electric fields perpendicular to the NV symmetry axis lift the remaining degeneracy by coupling the states $\ket{1}$ and $\ket{-1}$ at a strength of $d_\perp = 17\text{Hz~cm~V}^{-1}$.
Identifying $\ket{1}$ and $\ket{-1}$ as the detector qubit states ($X_1 = \ketbra{-1}{1} + \ketbra{1}{-1}$), this enables the measurement of a DC or AC electric field  
using standard Ramsey spectroscopy. In the case of AC measurements, a constant magnetic field has to be applied to bring the $\ket{1} \leftrightarrow \ket{-1}$ transition in resonance with the electric field frequency. 
 Due to the large zero-field splitting, $x$ and $y$ magnetic noise is highly suppressed [by a factor $(\omega_0 \tau_c)^2$], and the dominant noise contribution limiting the sensitivity is provided by magnetic field fluctuations in $z$ direction, accounting for pure dephasing of the qubit states. 
Furthermore, generically, the NV electron spin is hyperfine-coupled to the nuclear spin of the constituting nitrogen atom (whose coherence times are well beyond beyond those of the NV center \cite{Maurer:2012kg}), enabling coherent two-qubit operations \cite{Childress:2006km}. In particular, the longevity of the coupled nuclear spin has recently been used to successfully implement a full QEC procedure in two proof-of-principle experiments \cite{Waldherr:2013uq,Taminiau:2013up}.

For the simple QEC code we consider, the QEC operation $\mathcal R$ using the $^{15}$N nuclear spin as the ancilla can be done on the time scale of a few microseconds, without performing a full measurement and feedback loop, as described in \cite{SI}. 
This, in principle, allows extending the Ramsey time to the NV center population relaxation time $T\leq T_1$. Specifically, let us consider the case of DC or low frequency field sensing, relevant, e.g., in the biological imaging of neural activity \cite{Peterka:2011js,Robinson:2012fw,Kralj:2011hv}. In this case, DD cannot be used to improve the spin coherence time, and, generically, the interrogation time is limited by $T_2^*\approx 1 - 100\mu$s. Since depending on the operational conditions $T_1$ ranges from $10\text{ms}$ up to $1 \text{s}$, our QED approach could potentially improve the sensitivity by a factor of $\sqrt{T_1/T_2^*}=10-10^3$.
In the case of AC metrology, standard sensing experiments that use DD techniques such as Hahn echo or CPMG can achieve a suppression of the noise by a factor $(\Delta t / \tau_c)^2$ \cite{Viola:1999jw}, where $\Delta t$ denotes the duration of a single decoupling sequence.
Under typical experimental conditions this results in an effective coherence time of the order of $10 \mu\text{s} - 1\text{ms} \ll T_1$ (for shallow NV centers). In such experiments \cite{Dolde:2011hi}, QEC can still improve sensitivity by a factor of 3 to 300, reaching values of the order of $1-10 \text{V~cm}^{-1}\text{~Hz}^{-1/2}$ for a single NV center.

A second application of the QEC protocol involves AC magnetometry with NV centers. 
Alternatively to the conventional approach employing decoupling or double resonance techniques \cite{Loretz:2013io}, we consider a scheme in which we tune the transition frequency between the $\ket 0$ and $\ket{1}$ sublevels of the NV center ground state into resonance with the target AC field by applying an external magnetic field. As before, the use of a simple QEC protocol enhances the qubit coherence ideally to a value $\sim T_1$. 
As shown in \cite{SI}, similar to the above case, this approach can improve the sensitivity by a factor of 10 to $10^3$, and allows to expand the operational frequency range to several GHz. For applications requiring the use of diamond nano-crystals the improvement could, in principle, be markedly higher due to the lower initial spin coherence times.
The above considerations include the possibility of bulk magnetic and electric sensing with a macroscopic number of uncorrelated NV center spin detectors in a sample, since the QEC operation does not require individual addressing or measurement of different detector spins.  


In summary, we have presented a QEC-based approach to enhance the sensing accuracy in quantum metrology in the presence of noise. We demonstrated that our technique can improve the sensitivity of nanoscale magnetic and electric field sensors under current experimental conditions. Identifying further relevant physical situations in which QEC can be employed to improve sensing -- possibly by using more involved codes based on multiple qubits or multilevel systems -- remains an interesting task. In particular, the combination of the complementary techniques of QEC and DD in sensing protocols appears to be a promising path with potential applications in a large variety of fields \cite{Maze:2008ch,Taylor2008,London:2013fq,Loretz:2013io}.
From a theoretical perspective our approach demonstrates that sequential feedback protocols can improve the sensitivity bounds developed in \cite{Koiodynski:2013bu,DemkowiczDobrzanski:2012gl,Anonymous:GuhK9eRJ}. While the conditions we derived for perfect noise cancellation with QEC are restrictive, and applicable only to specific models, it remains an interesting questions if more general feedback protocols can be applied to more generic scenarios possibly at the cost of imperfect noise suppression.


\textit{Acknowledgements.} We thank Alex Retzker, Janek Kolodynski, and Luis Davidovich for enlightening
discussions.
This work was supported by NSF, CUA, HQOC, ITAMP, the Defense Advanced Research Projects
Agency (QuASAR program), and NDSEG (IL).

\appendix 
\section{Supplementary Information}
\subsection{A: Error analysis}

In this section we provide a full quantum mechanical derivation of the achievable measurement accuracy using QEC in the model described by Eq.~(4) of the main text.
For a given number of QEC steps $r$ we demand that $p_r \equiv \alpha \gamma \ll 1 $, i.e. the probability that two errors occur during the free evolution time is negligible [of order $\mathcal O(p^2)$]. 
Under this condition, and the assumption of perfect QEC gates, the QEC operation can reliably correct for errors after each segment $\alpha=T/r$. 
Note, that the above condition is by a factor $r$ less stringent than the naive condition without QEC $T\gamma\ll1$, and enables the extension of the interrogation time by the same factor.
Nevertheless, errors introduce a small residual phase uncertainty, due to the fact that the exact time of the error within the interval $\alpha$ is unknown. 
While in the absence of an error, the state picks up a phase $ \phi_0 = \omega \alpha$, this phase is reduced to a 
value $\phi(t) = \phi_0 - 2 \omega (\alpha-t)$, where $t\in (0,\alpha]$ denotes the time at which the error has occurred.
The probability density of errors occurring in $k$ segments at times $(t_1, t_2, \hdots, t_k),~ t_i\in [0,\alpha]$ is given by a binomial distribution 
\begin{align}
P_r(k) = {r \choose k} (p_r/\alpha)^k (1-p_r)^{r-k}.
\end{align}
For such an event, the total phase picked up by the state is
\begin{align}
\Phi_k(t_1,t_2,\hdots, t_k) = \Phi_0 - \sum_i^k 2\omega(\alpha-t_i),
\end{align}
where we defined $\Phi_0 = r \phi_0 = \omega T$.
Consequently, the probability that a given phase $\Phi$ has accumulated after the interrogation time $T$ is given by the distribution
\begin{align}
\xi(\Phi) = \sum_{k=0}^r P_r(k) \int_0^\alpha \hdots \int_0^\alpha dt_1'\hdots dt_k' \delta(\Phi_{k} - \Phi),
\end{align}
with $\delta(x)$ denoting the Dirac delta function. On average the state picks up the phase
\begin{align}\nonumber
\E \Phi =& \int d\Phi \xi(\Phi)\Phi \\ \nonumber
=& \sum_{k=0}^r P_r(k) \int_0^\alpha \hdots \int_0^\alpha dt_1'\hdots dt_k'  \Phi_k(t_1,\hdots,t_k)\\\nonumber
=&\sum_{k=0}^r P_r(k) (\alpha^k \Phi_0 - \alpha^{k+1} k \omega) \\
=& \Phi_0 - r p_r \omega \alpha = (1-p_r) \Phi_0.
\end{align}
This illustrated that instead of the actual frequency $\omega$, the presented measurement protocol rather assess the slightly modified parameter $(1-p_r)\omega$. This however does not present a limitation but merely requires an initial calibration of the device prior to the actual measurement. 

In an analogous but more involved calculation on further finds
\begin{align}
\E {\Phi^2} =  \left[(1-p_r)^2  + \frac{1}{r} (\frac{3}{4} p_r - p_r^2)\right] \Phi_0^2\equiv f(p_r)^2 \Phi_0^2,
\end{align}
where $f(p_r) \approx 1$, for ${p_r\ll1}$.

One then readily shows that the state of the detector qubit after a final disentangling operation subsequent to the last QEC step (see Fig.~1 of the main text) is given as
\begin{align}
\rho_{\Phi_0} = \int_{-\infty}^\infty d\Phi \xi(\Phi) \ketbra{\Phi}{\Phi},
\end{align}
where 
\begin{align}
 \ket\Phi =& \left( e^{-i\Phi}\ket{ \downarrow} +e^{i\Phi}  \ket{\uparrow} \right)/\sqrt2 \\
 =& \text{cos} (\Phi) \ket{+} + \text{sin}(\Phi) \ket{-},
 \end{align} 
with the $X$ basis states $\ket\pm =  (\ket{\uparrow}\pm \ket{\downarrow})/\sqrt 2$. Note, that the dependence of the state on the phase $\Phi_0$ that we want to measure is hidden in the distribution $\xi(\Phi)$.

A subsequent measurement in the $X$ basis yields the outcome $"1"$ with probability 
\begin{align}
P_+(\Phi_0) =& \text{Tr} \left(\ketbra{+}{+} \rho_{\Phi_0} \right)\\
=& \int d\Phi \xi(\Phi)\text{cos}^2\Phi\\ 
\approx& \frac{1}{2}\left(1+\text{cos}\left[2\sqrt{\E{\Phi^2}}\right]\right)\\
 =& \frac{1}{2}\left(1+\text{cos}\left[2f(p_r)\Phi_0\right]\right),
 \end{align} 
where in the last step we used the self-consistent assumption $\sqrt{\E{\Phi^2}}\ll1$. 

If we repeat this procedure $n$ times, this enables an estimation of the phase $\Phi_0$ with uncertainty
\begin{align}
\delta\Phi =& \frac{\sqrt{P_+(\Phi_0)[1-P_+(\Phi_0)]}}{\left| \partial P_+(\Phi_0)/ \partial \Phi_0 \right|} \frac{1}{\sqrt n}\\
=& \frac{1}{f(p_r)\sqrt n} \approx \frac{1}{\sqrt n}.
 \end{align} 

At the same time, since we only have to ensure that $p_r =T\gamma/r  \ll 1$, we can extend the interrogation time by a factor of $r$ as compared to the naive approach ${T \rightarrow r \gamma^{-1}}$.
By this we achieve a measurement accuracy
\begin{align}
\delta \omega \approx \delta \Phi /T \approx \sqrt \frac{\gamma}{r\tau},
\end{align}
$\sqrt r$ times better than the standard Ramsey protocol without QEC. Consequently, using $r= \tau/\gamma$ steps, we achieve the best accuracy allowed by quantum mechanics $\delta \omega \approx  {1}/{\tau}$.
Hereby the QEC operation is required to be fast on the timescale set by $\alpha \ll 1/\gamma$.

\subsection{B: Spontaneous emission and parallel dephasing}

In the main text we have demonstrated that if the signal generator and the noise operation elements fulfill conditions $(1) - (3)$, then QEC can successfully be applied to eliminate the effects of noise. Of course, for a given signal Hamiltonian there are certain types of noise that cannot be corrected, and the abovementioned conditions provide a useful tool to identify the latter.  

As an example, consider the case of a single qubit subject to spontaneous decay in the $z$ basis, accumulating a signal according to the Hamiltonian $H=\frac{\omega}{2} Z$ (an analogous argument can be given for an orthogonal signal $ H\propto X$). The error operation elements of spontaneous emission (with probability $p$) read
\begin{align}
E_{0}  = 
\begin{pmatrix} 
  1     & 0\\ 
  0 &  \sqrt{1-p} 
\end{pmatrix} ;
E_{1}=
\begin{pmatrix} 
  0     & \sqrt p\\ 
  0 &  0 
\end{pmatrix},
\end{align}
such that one readily shows $\mathcal G = Z \propto E_0^\dagger E_0 + \text{const}$. Since according to condition (1) $[\mathcal G, P]=0$ it directly follows that  $PE_0^\dagger E_0P= E_0^\dagger E_0 P$, violating condition (2), unless $P $ is an projector on a $Z$ eigenstate, in which case, however, $\xi =0$. Therefore any QEC code successfully correcting for independent spontaneous decay is useless for metrology. The same argumentation applies for pure dephasing noise in parallel direction of the signal ($H=\frac{\omega}{2} Z$, and $E_{1}\propto Z$).

\begin{figure}[!]
\centering
\includegraphics[width=0.45\textwidth]{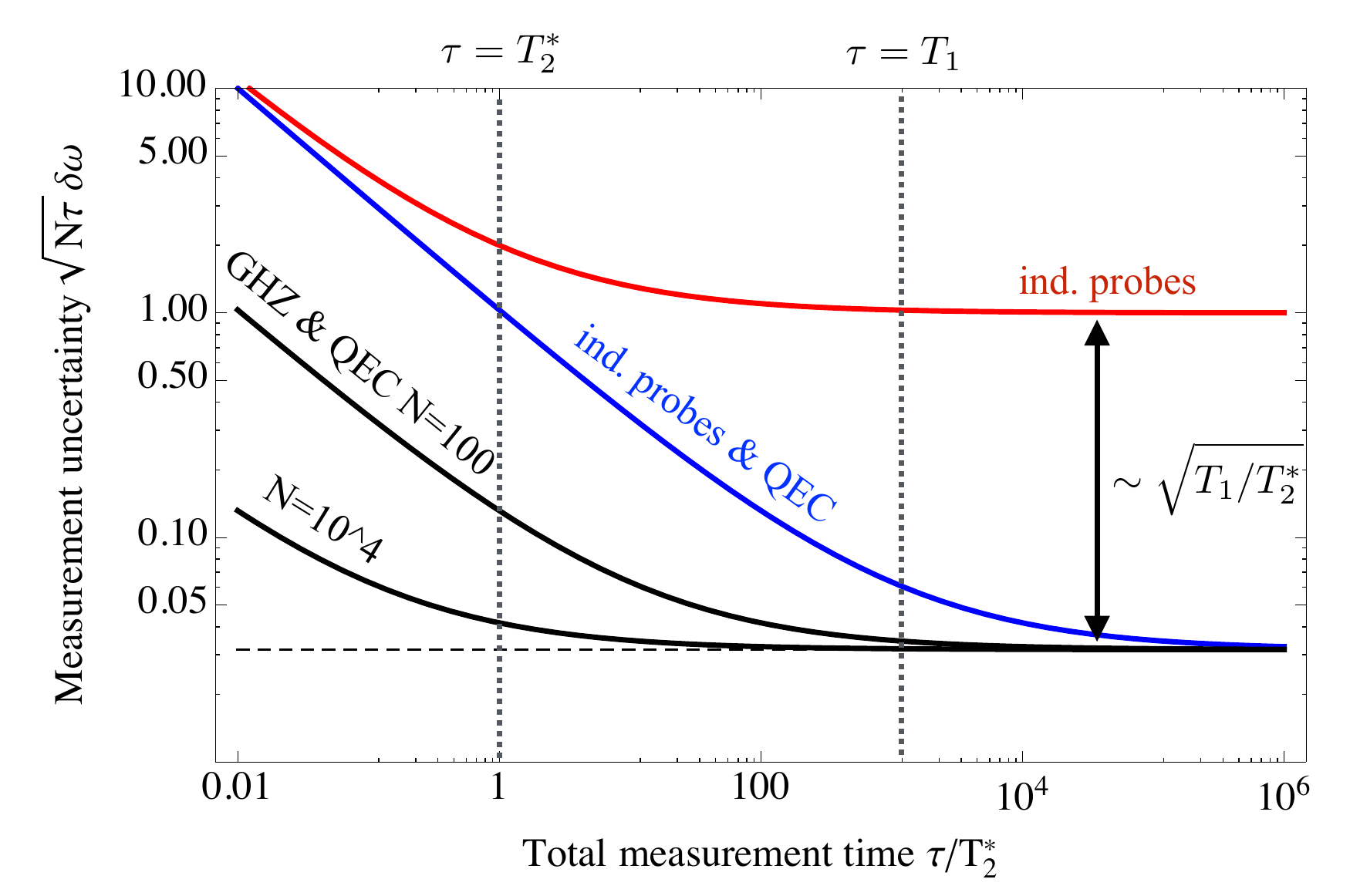}
\caption{Estimation error $\delta\omega$ (normalized by the total measurement time $\tau$, and the number of probe copies $N$) in electric field sensing with NV centers (see text) for $T_1=10^3T_2^*$. Independent probes in a standard Ramsey experiment without QEC (red line) are limited by the perpendicular magnetic noise giving rise to the electron coherence time $T_2^*$. If QEC if performed on each of the probes individually (blue line) the noise-free scaling can be extended up to the point where parallel noise components (in the form of $T_1$ relaxation processes) become relevant ($\tau \gtrsim T_1$), yielding a sizable constant improvement in the asymptotic limit. Using the multi-qubit code employing Greenberger-Horne-Zeilinger (GHZ) states as described in the main text (black lines), markedly larger improvement can be found in the limit of short measurement times. The parallel noise component leads to a breakdown of the Heisenberg scaling for $\tau \gtrsim T_1/N$. 
\label{fig:SI} }
\end{figure}

It should therefore be emphasized that the predicted noise-free scaling in the models considered in the main text strictly holds only in the regime where the inevitable uncorrectable noise components are negligible. 
Consider again the example of electric field sensing with NV centers. 
In a standard Ramsey-type experiment without QEC the interrogation time is limited by the perpendicular magnetic noise giving rise to the electron coherence time $T_2^*$. Therefore, for measurement times $\tau\gtrsim T_2^*$ the linear scaling of the sensitivity $\delta\omega\sim 1/\tau$ breaks down, and the suboptimal scaling  $\delta\omega\sim 1/\sqrt{T_2^*\tau}$ is found (red line in \figref{fig:SI}).
We have seen that this perpendicular noise component ($T_2^*$) can be corrected in a simple two qubit code allowing, in principle, to extend the interrogation time $T\rightarrow \tau$ as in a noise free situation. 
However, the population relaxation arising from the vibrational coupling to the diamond lattice, give rises to the NV centers spin relaxation time $T_1$. In a similar but slightly more complicated calculation than the one above one can show that such spin-flip errors are not correctable as they contain a parallel noise component. Therefore, for measurement times $\tau \gtrsim T_1$, the interrogation time will be limited by $T\leq T_1$ (blue line in \figref{fig:SI}).
In the asymptotic limit $\tau\rightarrow \infty$, this weaker restriction to the interrogation time yields a constant (and potentially large) sensitivity improvement by a factor $\sqrt{T_1/T_2^*}$. Note that if we have $N$ independent copies of the sensing qubit, and we perform QEC on each system individually, the sensitivity is improved by a factor $\sqrt{N}$. 

Let us now consider the multi-qubit QEC code described in the main text. There we demonstrated that provided with $N$ detector spins one can define the code $\ket{1}\equiv \ket{+}^{\otimes{N}}$ and  $\ket{0}\equiv \ket{-}^{\otimes N}$ to achieve Heisenberg scaling of the sensitivity, $\delta \omega \approx 1/(N\tau)$. Despite the fact that the parallel noise component (in the form of $T_1$ processes) leads to a breakdown of the Heisenberg scaling for $\tau\gtrsim T_1/N$ (reflected in the merging of the black solid and dashed lines for large $\tau$ in \figref{fig:SI}), for short measurement times $\tau$ the sensitivity improvements can be markedly higher ($\propto \sqrt{N}$) than for uncorrelated probes.

\subsection{C: QEC operation $\mathcal R$ for NV centers}
In this section, we discuss how the QEC operation $\mathcal R$ can be implemented for the example of  NV center  field sensing, using the $^{15}$N nuclear spin as an ancilla. In the first step, using standard techniques {\cite{Dutt:2007jn}}, we prepare state $(|0\downarrow\rangle + |1\uparrow\rangle)/\sqrt{2}$, where $|0\rangle$ and $|1\rangle$ denote the electron spin states, and $|\downarrow\rangle$ and $|\uparrow\rangle$ denote the projection of the nuclear spin on the NV axis. Applying a $\pi/2$ pulse to both the electron and the nuclear spins prepares the code state 
\begin{align}
\ket \Psi =(|++\rangle + |--\rangle)/\sqrt{2},
\end{align}
 which undergoes evolution, as described in the main text. The QEC procedure $\mathcal R$ is then implemented as follows. We again apply a $\pi/2$ pulse to both the electron and the nuclear spins. If no Z-error has occurred, the resulting state is 
\begin{align} 
\label{eq:1}
\ket {\Psi_\text{NE}} = (|0\downarrow\rangle + e^{i\Phi}|1\uparrow\rangle)/\sqrt{2},
 \end{align}
 whereas if a Z-error has occurred, the state is 
 \begin{align}
\ket {\Psi_\text{E}} = (|1\downarrow\rangle + e^{i\Phi}|0\uparrow\rangle)/\sqrt{2}.
\end{align} 
The error correction step itself consists of a CNOT gate that applies an electron $\pi$-pulse conditional on the $|\uparrow\rangle$ nuclear spin state (this is done by tuning the microwave frequency to the $|0\uparrow\rangle \rightarrow |1\uparrow\rangle$ transition). This operation decouples the electronic from the nuclear spin, storing the accumulated phase in the nuclear degree of freedom.
A subsequent pulse of green laser light \cite{Tamarat:2008ko} re-sets the electronic state [$\ket 1$ ($\ket{0}$) in the case of an (no) error] to $\ket{0}$, and another CNOT gate, as described above, recovers the state \eqref{eq:1}.
Application of a $\pi/2$ pulse, applied to both the electron and the nuclear spins, rotates the state back to the code space, completing the QEC operation. 
This procedure can be implemented on the time scale of a few microseconds: there is no need to perform a full measurement and feedback loop, due to the design of the re-setting operation using the polarizing optical transition of the NV center \cite{Tamarat:2008ko}.

\subsection{D: AC magnetometry with NV centers}

Using Ramsey-type measurements -- possibly improved by conventional DD techniques -- the sensitivity to classical AC magnetic fields is limited by the NV center coherence time, typically of the order of 10 $\mu$s to 1 ms, depending on the noise spectrum of the local environment. We consider an alternative scheme in which we tune the transition frequency between the $\ket 0$ and $\ket{1}$ sublevels of the NV center ground state into resonance with the target AC field by applying an external magnetic field which enables the measurement of fields with frequencies in a range from MHz to several GHz. By optically pumping the NV center into the $\ket 0$ sublevel and measuring the spin-dependent fluorescence after a variable delay time $T$, the spin flips induced by the target AC field can be monitored, constituting the magnetic signal. Ordinarily, the spin flips are suppressed by the broad linewidth ($T_2^* \sim 1$ to $100\mu$s) of the electronic transition, making this approach ineffective. However, performing QEC to effectively narrow the linewidth using the nitrogen nuclear spin as an ancilla qubit, we can suppress the dephasing due to the $z$ noise. The effectiveness of the QEC is independent of the environmental noise spectrum, and, as in the previous case, allows the Ramsey time to be extended to the phonon-induced $T_1$ time, improving the sensitivity by a factor of 10 to $10^3$, the frequency resolution by a factor of $10^2$ to $10^6$, and potentially expanding the operational frequency range to several GHz.


\begin{thebibliography}{36}
\bibitem[{fn()}]{fn11}
\bibinfo{journal}{We remark, that instead of the full feedback loop of measurement and correction, it is also possible to combine the two steps in a single operation, facilitating the experimental realization.}
\bibitem[{fn1()}]{fn1}
\bibinfo{journal}{This result is asymptotically true only for strictly perpendicular noise. A small noise component in parallel to the signal will eventually result in a breakdown of the asymptotic Heisenberg scaling. Nevertheless, as we demonstrate below, eliminating the dominant noise component results in a significant constant improvement, compare also \cite {SI}.}
\end{thebibliography}
\end{document}